# Novel Rattling of K Atoms in Aluminium-Doped Defect Pyrochlore Tungstate

Elvis Shoko[1a)], Gordon J Kearley[1], Vanessa K Peterson[1], Hannu Mutka[2], Michael M Koza[2], Jun-ichi Yamaura[3,4], Zenji Hiroi[3], Gordon J Thorogood[1]

1. Australian Nuclear Science and Technology Organisation, Locked Bag 2001, Kirrawee DC, NSW 2232, Australia
2. Institut Laue-Langevin, 6 Rue Jules Horowitz, B.P. 156, 38042 Grenoble Cedex 9. France.
3. Institute for Solid State Physics, University of Tokyo, Kashiwanoha, Kashiwa, Chiba 277-8581, Japan.
4. Materials Research Center for Element Strategy, Tokyo Institute of Technology, Nagatsuta 4259 S2-13, Yokohama, Kanagawa 226-8503, Japan.

## Abstract

Rattling dynamics have been identified as fundamental to superconductivity in defect pyrochlore osmates and aluminium vanadium intermetallics, as well as low thermal conductivity in clathrates and filled skutterudites. Combining inelastic neutron scattering (INS) measurements and *ab initio* molecular dynamics (MD) simulations, we use a new approach to investigate rattling in the Al-doped defect pyrochlore tungstates: $AAl_{0.33}W_{1.67}O_6$ (A = K, Rb, Cs). We find that although all the alkali metals rattle, the rattling of the K atoms is unique, not only among the tungstates but also among the analogous defect osmates, $KOs_2O_6$ and $RbOs_2O_6$. Detailed analysis of the MD trajectories reveals that two unique features set the K dynamics apart from the rest, namely, (1) quasi one-dimensional local diffusion within a cage, and (2) vibration at a range of frequencies. The local diffusion is driven by strongly anharmonic local potentials around the K atoms exhibiting a double-well profile in the direction of maximum displacement, which is also the direction of local diffusion. On the other hand, vibration at a range of frequencies is a consequence of the strong anisotropy in the local potentials around the K atoms as revealed by 'directional' magnitude spectra. We present evidence to show that it is the smaller size rather than the lighter mass of the K rattler which leads to the unusual dynamics. Finally, we suggest that the occurrence of local diffusion and vibration at a range of frequencies

---

a) Author to whom correspondence should be addressed. Electronic mail: elvis.shoko@gmail.com. Telephone: +61-41-506-4823

in the dynamics of a single rattler, as found here for the K atoms, may open new possibilities for phonon engineering in thermoelectric materials.

PACS number(s): 63.20.dd, 63.20.dk, 63.20.Pw, 63.20.Ry, 71.15.Pd, 72.20.Pa

## I. INTRODUCTION

The physical phenomenon termed "rattling dynamics"[1, 2] is not yet well understood and, consequently, no consensus exists on the use of this term. In this work, we define rattling as a local, low-frequency, and anharmonic thermal vibration of a guest atom that is weakly bound to its surrounding atoms that form an oversized atomic cage. Rattling dynamics have been identified as important to the superconductivity of defect pyrochlore osmates,[3, 4] aluminium vanadium intermetallics,[5-7] as well as the low thermal conductivity of clathrates[8] and filled skutterudites.[9-12] Rattling in the defect pyrochlore structure has been studied extensively in the osmates ($AOs_2O_6$, A = K, Rb, Cs) using various approaches.[13-25] In this study we report measured INS data and *ab initio* molecular dynamics (MD) simulations for the Al-doped tungstate pyrochlores, $AAl_{0.33}W_{1.67}O_6$ (A = K, Rb, and Cs) which are structurally analogous to the osmates (space group *Fd-3m*). The key structural feature of these materials is illustrated in FIG. 1 where the A cation (rattler) resides in a large cage formed by the W, Al, and O atoms, henceforth 'lattice atoms'.

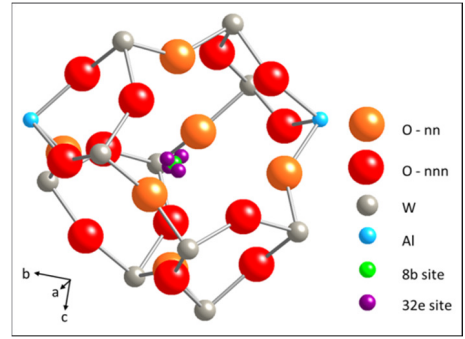

FIG. 1. (Color online) The tungstate cage consists of six nearest neighbor O atoms (O – nn) octahedrally arranged around the cage center followed by twelve next-nearest neighbor O atoms (O – nnn). Two W lattice sites per cage are doped with Al. The 'A' cation can occupy either of two sites: the 8*b* site at the cage center, and the 32*e* site consisting of four symmetry-equivalent positions tetrahedrally arranged around the cage center. The coordinate axes refer to the lattice vectors of the tetragonal supercell of the MD simulations (see FIG. 2).

## II. EXPERIMENT AND SIMULATIONS

The synthesis and characterization of the samples were described previously.[26] As per this earlier work, the compounds have space group *Fd-3m* with atomic positions: (0, 0, 0) for Al and W at the 16*c* site, (x, 1/8, 1/8) for O at the 48*f* site, and (x, x, x) site for the alkali metals at the 32*e* site.

We collected INS data using the IN6 instrument at the Institut Laue-Langevin (ILL), Grenoble, France. An incident wavelength of 4.14 Å was selected to give the best balance between energy range and resolution, and with the range of scattering angles, 10 to 115°, a momentum-transfer (Q) range of 0.26 to 2.55 Å$^{-1}$ was obtained. The experiment was performed as described by Mutka *et al.*[17] at 50 K for the K pyrochlore and 100 K for both Rb and Cs.

The *ab initio* MD simulations were performed using a tetragonal *3 x 1 x 1* supercell (216 atoms) consisting of 8 Al atoms substituting a selection of the 48 W atoms within the cell. This simulation supercell, chosen to achieve the desired stoichiometry, is illustrated in FIG. 2. As there is no unique way of making the selection of Al substitution for W, the choice was guided by the likely scenario where the Al atoms are well separated from each other. The Projector-Augmented Wave (PAW) method implemented[27, 28] in the Vienna Ab-initio Simulation Package (VASP)[29, 30] was employed, and for the exchange-correlation potential, the generalized gradient approximation with the Perdew, Burke, and Ernzerhof (GGA-PBE) functional was used.[31, 32]

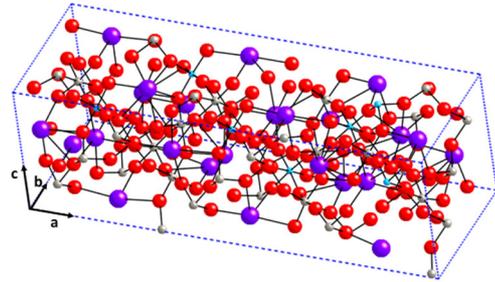

FIG. 2. (Color online) MD simulation supercell (3 x 1 x 1) consisting of 216 atoms as follows: 24 alkali metal (violet), 40 W (grey), 8 Al (light blue) and 144 O atoms (red). The cell axes are as labeled.

All structures were first relaxed to obtain optimized geometries within the experimental lattice parameters before equilibration at 100 K. All simulations were then run at this temperature, which matches that of the INS experiment for Rb and Cs, but differs from the 50 K used in the K experiment. For both the K and Rb simulations, a microcanonical ensemble (NVE) was implemented, although both systems cooled slowly (due to the inherent round errors of the numerical algorithms) so that for the duration of the simulation, the temperature fluctuated within the 100 - 60 K range. This cooling necessitated an additional simulation at 150 K for the K which was more stable permitting a longer simulation to get reliable statistics for single-site analysis.

We were unable to establish a stable NVE simulation for the Cs system because of an unrealistic rapid increase in the temperature after only a few picoseconds. For this reason the Cs system was simulated in the canonical ensemble (NVT), where a constant temperature was achieved through a simple velocity scaling scheme in VASP. Γ-point sampling was performed, and all simulations were a minimum of 24 ps with a time step of 1 fs. The MD spectra were calculated using nMOLDYN[33] with a Gaussian width of 0.3 meV selected to correspond to the instrument resolution in the INS experiment, while magnitude spectra were obtained from the (forward) fast Fourier transform (FFT) of the MD trajectories. Details of the simulation settings for the osmium analogues, i.e., $KOs_2O_6$ and $RbOs_2O_6$ are reported elsewhere.[34]

## III. RESULTS AND DISCUSSION

Our method for assessing the rattling characteristics of a material involves calculating simulated elemental spectra, atomic root-mean-square displacements (rmsds), and Pearson coefficients to determine atomic dynamical correlations. Where appropriate, analysis of the vibrational dynamics is performed in the principal axis coordinates of the motion of the atoms[35]. As a tool for assessing the nature of the diffusive motion, we calculate the mean-square displacement autocorrelation function (msd):[36, 37]

$$msd(j) = \frac{1}{N-j} \sum_{i=1}^{N-j} \left[ \mathbf{R}(t_i) - \mathbf{R}(t_{i+j}) \right]^2, \quad j = 1, 2, ..., N-1 \qquad (1)$$

Where $\mathbf{R}(t_i)$ and $\mathbf{R}(t_{i+j})$ are the nuclear coordinates at the $t_i$ th and $t_{i+j}$ th simulation time steps, respectively, and $N$ is the total number of time steps in the simulation. In this work, the msd is calculated as an average over all the 24 alkali-metal atoms in the simulation cell for the doped compounds, and 8 for the undoped. The physical meaning of the msd function is understood by considering two simple limiting cases of the dynamics of an atom. In the first case, the atom undergoes long-range diffusion in such a way that its distance from the starting point is always increasing. In this case, the msd is a monotonically-increasing function of time. In the second case, the atom vibrates harmonically about its mean position so that its displacement in one dimension (1D) (e.g., along $x$) can be written as: $x(t) = A*\sin(\omega t + \varphi)$, where $A$ is the amplitude, $\omega$ is the frequency (rads/s), and $\varphi$ is the phase angle. For this case, the msd function oscillates between 0 and $2*A^2$ with frequency $\omega$. Real dynamics of atoms fall somewhere between these extremes; on the one hand, being closer to the monotonically-increasing function if they undergo long-range Fickian diffusion, and on the other, being closer to the uniformly-oscillating function if the local potential is approximately harmonic.

We study the local potentials around the K atoms at different temperatures by calculating the potential of mean force (PMF):[38]

$$PMF = -k_B T * \log P(\mathbf{r}) \qquad (2)$$

With $k_B$, the Boltzmann constant, $T$, absolute temperature, and $P(\mathbf{r})$ the population density at position $\mathbf{r}$ for a given atom. In one dimension (1D), the PMF is the free energy curve along a specified direction and consists of the potential energy part and the entropy. Here, we calculate a PMF along the direction of maximum displacement for the alkali-metal atoms as an average over all these atoms in the simulation cell. Each atom is centered at its time-average position and displacements are then calculated relative to this position. The full range of the displacements for all the atoms is then partitioned into suitable intervals from which a histogram of the displacements is constructed for all the atoms over the duration of the simulation. The population density, $P(\mathbf{r})$ in Eq. (2), is calculated from this histogram.

For discussions involving magnitude spectra, we are only interested in comparing the spectral forms, and therefore the magnitude spectra are suitably scaled to facilitate this type of analysis. Here, the absolute values of the magnitude in these spectra are not relevant as these depend on a number of variables (e.g., temperature and duration of simulation) which may be different between simulations.

## A. Validation of simulations

Before a detailed analysis of the MD trajectories, we first perform a validation of the simulations by comparing to the experimental structures and dynamics.

### 1. Validation of MD average structure

Table I summarises the positional parameters of the alkali metal atoms calculated from the MD compared to those from experiment. The results show some discrepancy between the two which could arise from at least two sources: Firstly, the static disorder due to the Al doping means that there exists a vast number of possible configurations (~ $10^6$) and we only used one of these, in contrast to the experiment which measures the ensemble average of the complete set. As can be seen in Table I, the average structure calculated from an MD simulation of a K-pyrochlore obtained by simulated annealing of the original structure is also slightly different to that of the original structure. Secondly, for systems dominated by rattling dynamics which involve large displacements of the rattlers, to get an accurate average structure, the simulation needs to be run long enough for the rattler to explore all its available volume.

Table I. Positional parameters of the alkali metal atoms from experiment[26] and those calculated from the MD simulation. The main source of the differences is the static disorder in these compounds due to Al doping. The MD result is from only one of a vast number of configurations, whereas the experiment includes all configurations.

| Alkali metal | Method | $x$ | $y$ | $z$ |
|---|---|---|---|---|
|   | Exp. | 0.3445 | 0.3445 | 0.3445 |
| K | MD | 0.3121 | 0.3475 | 0.3471 |
|   | MD (annealed) | 0.2790 | 0.3240 | 0.3282 |
| Rb | Exp. | 0.3569 | 0.3569 | 0.3569 |
|   | MD | 0.3872 | 0.3592 | 0.3393 |
| Cs | Exp. | 0.3618 | 0.3618 | 0.3618 |
|   | MD | 0.3528 | 0.3628 | 0.3598 |

In this regard, we point out that, because INS is a local probe, the MD simulation may accurately capture local dynamics in the INS without necessarily accurately reproducing the experimental average structure. This is because the dynamics are similar despite the site-to-site variations in local structure due to the static disorder.

### 2. Validation of MD dynamics

In FIG. 3, the simulated INS spectra calculated from the MD results and the experimental INS spectra are compared and it is noted that the MD simulations reproduce the experimental spectra quite well. The negative energy indicates energy transfer from the sample to the probing neutrons in the experiment. The reasonable agreement between the MD and experiment obtained in FIG. 3 implies that the underlying MD trajectories are adequate to study the detailed dynamics of these systems.

## B. Evidence for alkali-metal rattling

Figure 4 shows the elemental contributions to the spectra of FIG. 3. These partial spectra reveal two important features of interest to this study. Firstly, FIG. 4 shows a striking dominance of alkali-metal modes at low energies. The peaks are relatively sharp, particularly for Rb and Cs, which evidences weak coupling between the alkali-metal vibrations and those of the rest of the lattice, a characteristic feature of rattling dynamics. Secondly, FIG. 4(a) shows that the quasielastic profile of the K-pyrochlore spectra evident in FIG. 3 arises from the dynamics of the K atoms. The distinction between the K dynamics and those of Rb and Cs is more clearly exhibited in the insets of FIG. 4; while the K spectrum is virtually structureless with a quasielastic profile, both Rb and Cs show, in contrast, exhibit well-localized peaks characteristic of oscillatory dynamics, FIG. 4 (b) and (c).

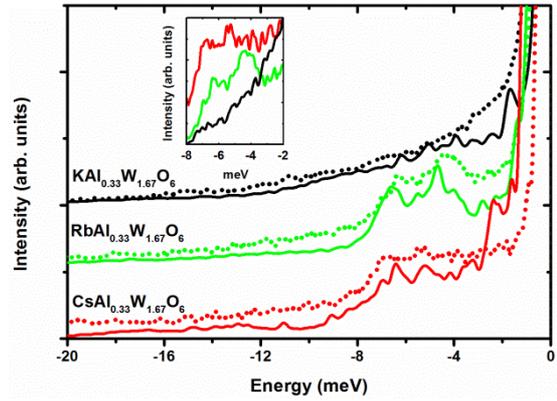

FIG. 3. (Color online) The measured INS spectra (dotted lines) are compared to the simulated spectra calculated from the MD simulations (continuous lines). For clarity, spectra are displaced along the arbitrary intensity axis and they are also suitably scaled to facilitate comparison. Reasonable agreement is obtained between the spectra in this energy range indicating that the MD simulation describes the physical dynamics well. The alkali-metal modes are low-energy, 2-8 meV, and are highlighted in the inset, where the spectra are displaced upwards in the order K, Rb, and Cs for clarity. The absence of any prominent peak features from the K spectrum is striking indicating dynamics quite distinct from both Rb and Cs.

Figure 4(a) also shows that the lattice modes of the K-pyrochlore exhibit low-energy features somewhat similar to those of the K modes. This similarity in spectral profiles may suggest that the K atoms are somewhat more sensitive to the lattice modes than Rb and Cs.

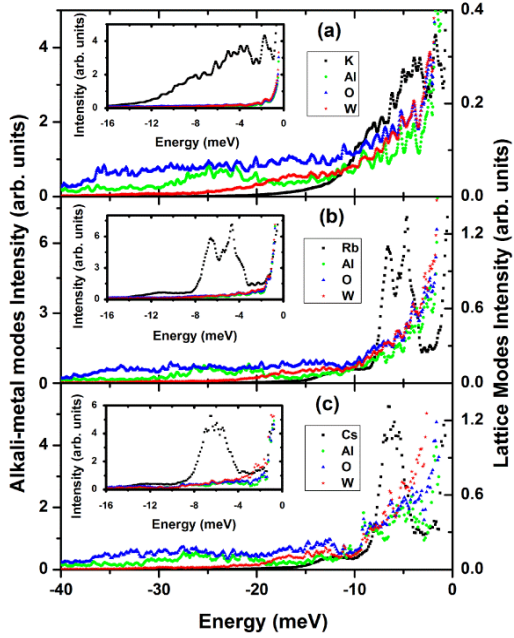

FIG. 4. (Color online) Elemental decompositions of the MD spectra of FIG. 3, showing the K, Rb, and Cs tungstates from the top to bottom panel, respectively. The alkali-metal modes are plotted on the left scales while the lattice atom modes are on the right scales. For all the pyrochlores, the alkali-metal modes dominate the low-energy modes as is highlighted in the respective insets by plotting all the spectra on the same scale.

To fully establish rattling dynamics we also need to show that the alkali-metal modes do not couple to the optical phonons of the lattice. One simple way to assess this is to calculate the (Pearson's) correlation between distances of suitably selected atoms of the lattice and the distance between the alkali metal and an atom of the lattice. We performed this type of calculation for the K-pyrochlore using the cluster illustrated in Figure 5. The resulting correlation matrix, Table II, shows that the relevant correlations (bold print) are low, indicating that only weak coupling exists between the K- and the optical modes.

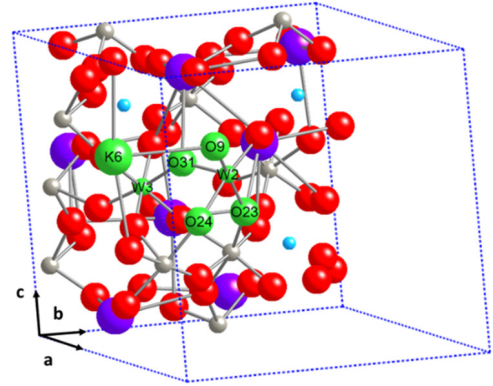

FIG. 5. (Color online) Part of the MD supercell to show the cluster of atoms (highlighted in green) used to assess the extent of coupling between the alkali-metal modes and the optical phonons of the lattice. The numbers refer to the atom labels in the supercell and for the O atoms, O9 is the nearest-neighbor to the K atom K6, O24 and O31 are the next-nearest-neighbors to K6, and O23 the next-next-nearest-neighbor to K6. Atoms not in the highlighted cluster are: K (violet), O (red), W (grey), and Al (light blue). The cell axes correspond to those shown in FIG. 1.

**Table II.** The correlation matrix calculated for the cluster shown in FIG. 5. The relevant Pearson's correlation coefficients for assessing the coupling between the K atom K6 and the lattice optical modes are highlighted in bold print, with atom distances labeled A-J for ease of discussion. The correlation for distances E-I is somewhat higher than expected but overall, the relevant correlations are low indicating weak coupling of the rattler to the lattice optical phonons.

|   | A: K6–W2 | B: K6–O9 | C: O9–W2 | D: W2–O24 | E: O24–K6 | F: W2–O23 | G: O31–O9 | H: W2–O31 | I: W3–W2 | J: K6–W3 |
|---|---|---|---|---|---|---|---|---|---|---|
| A: K6–W2 | 1.00 | 0.88 | −0.16 | 0.00 | 0.90 | **0.15** | **0.13** | −0.15 | 0.29 | 0.32 |
| B: K6–O9 | 0.88 | 1.00 | **−0.23** | **−0.05** | 0.72 | **0.16** | 0.23 | **−0.14** | **0.22** | 0.03 |
| C: O9–W2 | −0.16 | **−0.23** | 1.00 | 0.13 | **−0.20** | −0.59 | −0.02 | −0.06 | −0.19 | **−0.11** |
| D: W2–O24 | 0.00 | **−0.05** | 0.13 | 1.00 | −0.05 | −0.33 | −0.23 | −0.15 | 0.14 | **−0.16** |
| E: O24–K6 | 0.90 | 0.72 | **−0.20** | −0.05 | 1.00 | **0.24** | **0.12** | **−0.19** | **0.33** | 0.32 |
| F: W2–O23 | **0.15** | **0.16** | −0.59 | −0.33 | **0.24** | 1.00 | 0.23 | 0.03 | 0.28 | **0.21** |
| G: O31–O9 | **0.13** | 0.23 | −0.02 | −0.23 | **0.12** | 0.23 | 1.00 | 0.06 | 0.19 | **0.24** |
| H: W2–O31 | −0.15 | **−0.14** | −0.06 | −0.15 | **−0.19** | 0.03 | 0.06 | 1.00 | −0.02 | **−0.10** |
| I: W3–W2 | 0.29 | **0.22** | −0.19 | 0.14 | **0.33** | 0.28 | 0.19 | −0.02 | 1.00 | −0.04 |
| J: K6–W3 | 0.32 | 0.03 | **−0.11** | **−0.16** | 0.32 | **0.21** | **0.24** | **−0.10** | −0.04 | 1.00 |

Table III provides further evidence for alkali-metal rattling, where all the alkali-metal atoms exhibit relatively large rmsds and low correlations, consistent with weak bonding. The K atoms show large scatter indicating unusual dynamics involving local diffusion in a cage as will be discussed in Sec. C. 3.

**Table III.** Rmsds (Å) and Pearson correlation coefficients calculated from the MD for the K-, Rb-, and Cs-pyrochlores along the direction of maximum displacement. The rmsds are averages over all atoms of the same element while the correlation coefficients are calculated from the averages of the absolute values of the atom/neighbor correlations up to 3.5 Å. In column 2, A = K, Rb, or Cs. The results show an inverse relationship between the correlation coefficient and the rmsds

which is expected since an atom which is weakly bonded should move more freely. Note the large scatter in the K rmsd (last column) compared to Rb and Cs.

| Compound | Measure | A | W | Al | O | Range for A |
|---|---|---|---|---|---|---|
| $KAl_{0.33}W_{1.67}O_6$ | rmsd | 0.168 | 0.042 | 0.039 | 0.053 | 0.089-0.318 |
| | Corr. Coeff. | 0.082 | 0.328 | 0.306 | 0.265 | 0.039-0.142 |
| $RbAl_{0.33}W_{1.67}O_6$ | rmsd | 0.119 | 0.051 | 0.047 | 0.067 | 0.097-0.128 |
| | Corr. Coeff. | 0.070 | 0.310 | 0.288 | 0.257 | 0.028-0.139 |
| $CsAl_{0.33}W_{1.67}O_6$ | rmsd | 0.111 | 0.077 | 0.051 | 0.071 | 0.083-0.165 |
| | Corr. Coeff. | 0.091 | 0.315 | 0.291 | 0.256 | 0.026-0.179 |

### C. Novel rattling of K atoms

From now on, we exclude the Cs results because of the limitations of simple velocity scaling NVT. As already noted, the K system spectra are distinct in being virtually structureless with a quasielastic profile. From our experimental INS, we noted that the total scattering, elastic incoherent contribution, and the inelastic contribution of the K compound were about four, ten, and five times, respectively, that of the Rb and Cs counterparts, which led us to suspect the presence of residual $H_2O$ molecules which could cause a significant quasielastic signal. However, the good agreement between the experimental and MD spectra in FIG. 3 cannot be fortuitous, and this excludes that explanation for the spectral profile.

Figure 6 demonstrates that the quasielastic spectrum for the K of $KAl_{0.33}W_{1.67}O_6$ is distinct among similar compounds. The contrast between the K spectrum of $KAl_{0.33}W_{1.67}O_6$ and that of $KOs_2O_6$ is particularly noteworthy and reflects at least some effects of the Al-doping and the smaller rattler/cage radius ratio in the former (0.136) than in the latter (0.137).

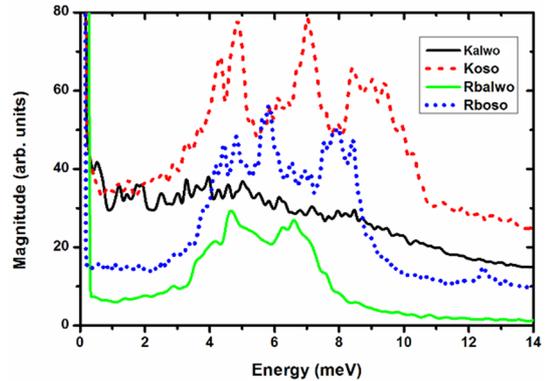

FIG. 6. (Color online) Simulated INS spectra calculated for different compounds from MD simulations at 100 K. The different spectra are: Kalwo - the K modes in $KAl_{0.33}W_{1.67}O_6$, Koso – K in $KOs_2O_6$, Rbalwo – Rb in $RbAl_{0.33}W_{1.67}O_6$, and Rboso – Rb in $RbOs_2O_6$. The spectra of all the other compounds exhibit oscillatory dynamics while that of $KAl_{0.33}W_{1.67}O_6$ has a distinctly quasielastic profile, reflecting unusual rattling dynamics. Magnitudes of **the spectra are suitably scaled and shifted to facilitate visual comparison.**

### 1. Starting configuration and rattler atomic mass

The static disorder from Al doping means that there are several possible different starting configurations for the simulation that may lead to different alkali-metal dynamics. To investigate this, we performed an MD simulation (100 K, 36 ps) on a configuration obtained from the simulated annealing to 500 K of our normal configuration. The results in FIG. 7 (K_annealed) show that these two different simulation starting configurations produce similar spectra which preserve the quasielastic profile. Thus the K dynamics are relatively insensitive to the starting configuration of the simulation. We also investigated whether the light mass of the K could contribute to its novel dynamics by performing MD simulations on two artificial systems in which the mass of the K atom was set to that of Rb and Cs. The calculated spectra in FIG. 7, labeled K_Rb (37 ps) and K_Cs (17 ps) show that the spectra of these 'heavier K' atoms preserve the quasielastic profile, indicating that ionic size is more important than mass. In the K, Rb, and Cs pyrochlore series, the ratios of the rattler's ionic radius to the lattice constant of the respective pyrochlores are: 0.136, 0.149, and 0.164, indicating a smaller ratio for the K.

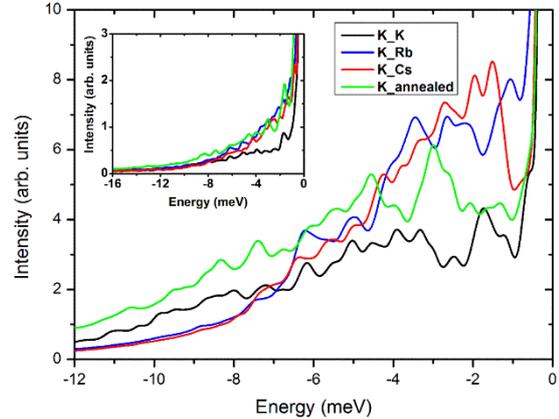

**FIG. 7. (Color online) Calculated elemental INS spectra from the MD simulation for the K-pyrochlore shown to illustrate the role of starting configuration and rattler mass in the unusual K dynamics. The spectra are labeled as follows: K_K and K_annealed are the normal K-pyrochlore starting configurations, with the latter annealed (to 500 K), while K_Rb, and K_Cs are the K-pyrochlore with the mass of all the K atoms set equal to Rb, and Cs, respectively. INSERT: The full calculated INS spectra. These results show that the K dynamics are relatively insensitive to both the starting configuration and the mass of the rattler.**

### 2. Local Potentials

To understand the origin of the novel K dynamics, we calculated 1D PMFs to get visual profiles of the average local free energy along selected directions. We first show in FIG. 8 that, within the same compound ($KAl_{0.33}W_{1.67}O_6$), the PMF analysis reveals the rigidity of the lattice and weak bonding for the alkali metals (K).

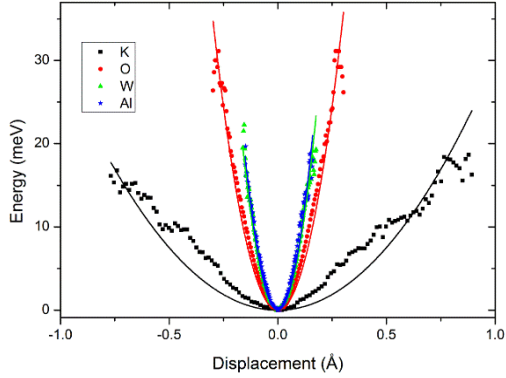

**FIG. 8.** (Color online) The PMFs of all the different atom types in the K-pyrochlore supercell plotted along the direction of maximum displacement at the mean temperature of 74 K. Continuous lines are fits to the harmonic oscillator model. Compared to the other atoms, the K PMF is broader and highly anharmonic as can be seen from the poor harmonic fits. The PMFs of the lattice atoms are significantly narrower and approximately harmonic indicating rigidity of the cage.

Force constants (i.e., a local force constant of the atom interacting with the entirety of its environment) extracted from the harmonic fits in FIG. 8 are 0.96, 12.52, 24.09, and 24.13 N/m, corresponding to vibrational frequencies of 2.5, 14.3, 5.8, and 15.3 meV, for K, O, W, and Al, respectively. Although theoretically these frequencies could be matched to the spectra, this will only be meaningful for the K spectrum, as the other spectra consist of many acoustic modes arising from complex motions in multiple directions. We note that although the force constants we calculate here are not the spring constants of the atom-atom bonds, they directly evidence how strongly an atom is bonded. Consequently, one concludes from FIG. 8 that the K atoms are weakly bonded to their local environment and the cage is rigid.

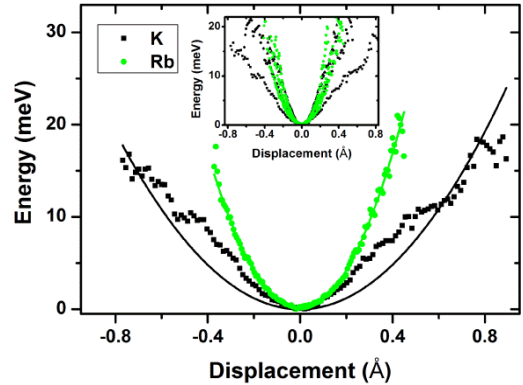

**FIG. 9.** (Color online) The 1D PMFs plotted along the direction of maximum displacement at the mean temperatures of 74 K and 77 K for K and Rb, respectively. Continuous lines are fits to the harmonic oscillator model. The K PMF in $KAl_{0.33}W_{1.67}O_6$ is considerably broader and highly anharmonic compared to that of Rb in $RbAl_{0.33}W_{1.67}O_6$. INSET: The K and Rb PMFs along the three principal axes are plotted to highlight the significant anisotropy in the K PMF.

Figure 9 shows high anisotropy and strong anharmonicity for the K PMF and we examine these two aspects to establish their influence on the K dynamics. We perform a site-by-site analysis to connect each site trajectory to the corresponding PMF. For reliable single-site statistics, a longer simulation (51 ps) at 150 K is used in place of the less thermally stable one at 100 K which cooled down to 74 K after only 27 ps. Figure 10(a) shows that the K dynamics at the two temperatures are similar which enables us to extract the more detailed information from the longer 150 K simulation.

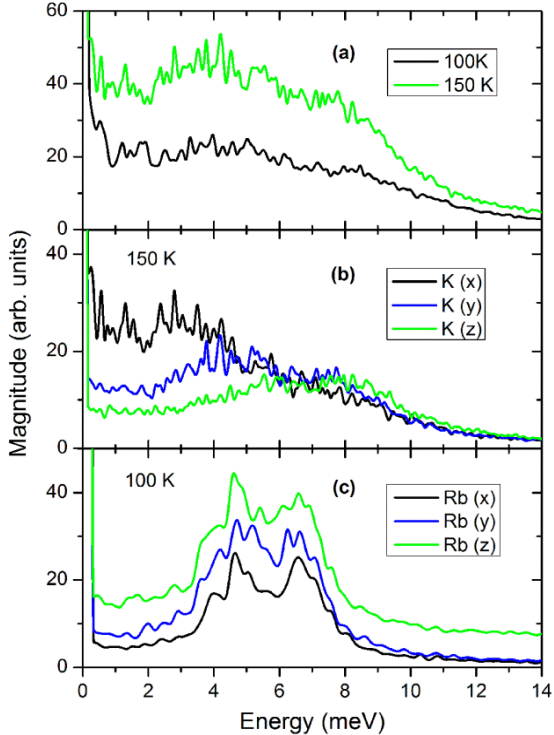

FIG. 10. (Color online) Alkali-metal magnitude spectra: (a) K in $KAl_{0.33}W_{1.67}O_6$ at 100 K and 150 K reflecting similar quasielastic spectral features, (b) and (c), partial spectra for K and Rb along the principal axes, with displacements decreasing in the order x, y, and z. Panel (b) shows that the anisotropy in the K PMF shifts frequencies to higher energy from x to z, with the dynamics becoming more oscillatory, whereas panel (c) shows virtually no anisotropic effects on the Rb spectra. Spectra are scaled and displaced along the magnitude axis to facilitate visual comparison.

Figure 10(b) shows strong directionality in the spectra of the K atoms, indicating that the strong anisotropy of the K PMFs causes significant phonon softening in the direction of maximum displacement leading to the K atoms vibrating at a much broader range of frequencies than Rb. In contrast, there is virtually no directionality for the Rb spectra in FIG. 10(c). Additionally; Figure 10(b) reveals that the quasielastic profile of the K spectrum only occurs in the direction of maximum displacement indicating that the K dynamics are, in fact, oscillatory in the other two directions. Thus the diffusive K dynamics are quasi-1D.

### 3. Local Diffusion

We found that the width of the K experimental spectrum as a function of Q is approximately constant indicating local motion. Since Fickian diffusion generally leads to a peak width varying as ~ $Q^2$, this is conclusive evidence that the K does not undergo long-range diffusion.[39] From single-site analyses, we now examine the exact nature of the unusual K dynamics in the direction of maximum displacement by selecting two sites representing the strongest and weakest quasielastic signals, respectively. The spectra of the two selected sites, K4 and K12, are shown in FIG. 11 (a) where K4 has the strongest quasielastic signal. The respective site PMFs in the direction of maximum displacement are plotted in FIG. 11(b), with continuous lines representing fits to the harmonic oscillator model. Both PMFs exhibit poor fits to the harmonic model with flatter profiles in the neighborhood of the time-average position, indicating significant anharmonicity. The stronger the quasielastic signal, the stronger is the anharmonicity with K4 showing a much broader profile exhibiting a double-well signature. The msd autocorrelation functions plotted in FIG. 11(b) inset show a clear difference between K4 and K12; K4 exhibits an aperiodic function that rises and falls over relatively long time durations indicative of local diffusion. In contrast, the function for K12 reflects a stronger oscillatory character with much smaller displacements indicative of the oscillatory dynamics already exhibited in FIG. 11(a). Thus, msd autocorrelation functions plotted in FIG. 11(b) inset, clearly show that the quasielastic profile of the K4 spectrum arises from the quasi-1D local diffusion confined to the direction of maximum displacement inside the cage.

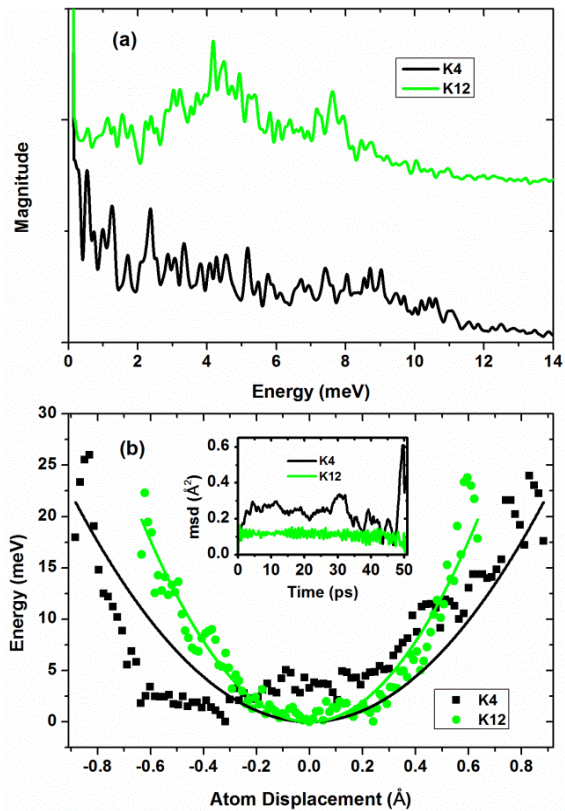

FIG. 11. (Color online) (a) Magnitude spectra for two selected alkali-metal sites (K4 and K12) from the 150 K simulation of $KAl_{0.33}W_{1.67}O_6$, and (b) the corresponding PMFs and msds (inset). The magnitude spectra are displaced along the magnitude axis to facilitate visual comparison. Panel (b) shows that the quasielastic signal for K4 (panel (a)) is a consequence of a highly anharmonic and approximately double-well potential in which the atom undergoes local diffusion between the wells as evidenced by the profile of its msd (inset).

We examined the possibility that the K could be hoping between two 32*e* sites, an issue which appears to still be in dispute[40-42] for the $KOs_2O_6$ case. Figure 12 shows that the K is (0.33 Å) off-center, and that the two wells in the PMF do not coincide with the 32*e* sites from Galati *et al*[43]. The MD evidence for off-center location is strong but a complete discussion of the site-to-site hopping requires a longer simulation.

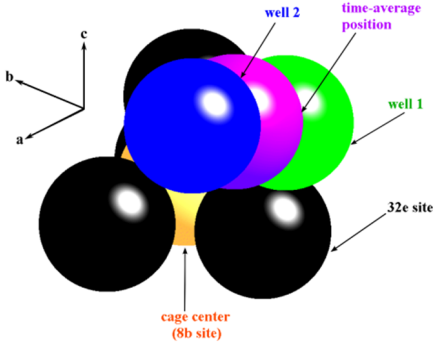

**FIG. 12. Schematic of the 8*b* and 32*e* sites, time-average position, and the PMF wells for K4. The 32*e* sites coordinates[43] are corrected for the calculated MD cage center and the displacement ranges are -0.585 - -0.298 Å and 0.038 – 0.115 Å for well 1 and well 2, respectively (see FIG. 11(b)). The wells are 0.32 Å apart (cf. 0.46 Å between 32*e* sites) and lie on a line approximately perpendicular to the line connecting the closest 32*e* site pair. The coordinate axes refer to the lattice vectors of the tetragonal supercell of the MD simulations (see FIG. 2).**

Although the existence of a direction in which the K undergoes large anharmonic excursions driven by a quasi-1D double-well potential is somewhat similar to what was reported for $KOs_2O_6$ along the (111) direction[19, 20, 44], any comparisons require caution as FIG. 6 clearly indicates that the dynamics are different between these compounds. The combination of local diffusion and vibration at a range of frequencies in the dynamics of a single rattler as exhibited by the K atoms in $KAl_{0.33}W_{1.67}O_6$ could be more effective at scattering the heat-carrying acoustic phonons compared to a rattler exhibiting only one of these properties. Thus the novel K dynamics elucidated in this work could open a new approach for reducing the lattice thermal conductivity in thermoelectric cage compounds.

## IV. CONCLUSION

In conclusion, we have shown that all the alkali metal atoms (K, Rb, and Cs) in the Al-doped defect pyrochlore tungstates rattle with the K exhibiting novel dynamics involving quasi-1D local diffusion inside the cage as well as vibrating at a broad range of frequencies. These effects, driven by highly anisotropic and anharmonic potentials are a consequence of the smaller size of the K atom relative to its cage volume. Overall, the insights into the microscopic picture of rattling dynamics, particularly the novel K dynamics, along with our method of characterising them may have broader implications for similar materials. For instance, our method could be used to clarify how rattlers moderate the lattice thermal conductivity in clathrates, skutterudites, and other cage compounds being investigated for thermoelectric applications.

## ACKNOWLEDGEMENTS

We gratefully acknowledge helpful discussions with Dr. R. Kutteh on various technical issues regarding VASP simulations.